# Flexoelectricity induced increase of critical thickness in epitaxial ferroelectric thin films

Hao Zhou [a], Jiawang Hong [b], Yihui Zhang [b], Faxin Li [a], Yongmao Pei [a,*], and Daining Fang [b,*]

[a] *State Key Laboratory for Turbulence and Complex Systems, College of Engineering, Peking University, Beijing 100871, China*

[b] *Department of Engineering Mechanics, Tsinghua University, Beijing 100084, China*

**ABSTRACT**

Flexoelectricity describes the coupling between polarization and strain/stress gradients in insulating crystals. In this paper, using the Landau-Ginsburg-Devonshire phenomenological approach, we found that flexoelectricity could increase the theoretical critical thickness in epitaxial $BaTiO_3$ thin films, below which the switchable spontaneous polarization vanishes. This increase is remarkable in tensile films while trivial in compressive films due to the electrostriction caused decrease of potential barrier, which can be easily destroyed by the flexoelectricity, between the ferroelectric state and the paraelectric state in tensile films. In addition, the films are still in a uni-polar state even below the critical thickness due to the flexoelectric effect.

*Corresponding authors, E-mail addresses: peiym@pku.edu.cn (Yongmao Pei); fangdn@pku.edu.cn (Daining Fang)

## 1. Introduction

Flexoelectricity (FxE) describes the coupling between polarization and strain/stress gradients in insulating crystals. Although it has been discovered for more than 40 years [1], FxE did not arouse much attention for a long time, since it is not obvious in bulk materials. With the development of synthesis of ferroelectric films widely used in microelectronic devices, FxE, which can exist in all dielectrics under inhomogeneous deformation, has raised great concern in recent years, because it may significantly affect the functional properties of films, superlattices and nanostructures. For instance, it may be responsible for a variety of anomalous phenomena, such as the imprint behavior of ferroelectric thin films [2-4] and the "dead layer" effects in nanocapacitor systems [5].

The historical development of FxE study has been reviewed by Tagantsev [6] and more recently by Cross [7]. A series of experiments have been done by Ma and Cross [8] and large FxE coefficients in several ferroelectric ceramics have been found. Zubko *et al.* obtained the full FxE tensor in the single crystals of the paraelectric $SrTiO_3$ [9]. Catalan *et al.* also found the FxE coupling played a fundamental role in the dielectric constant of epitaxial ferroelectrics thin films [10-11]. Recently, Shen et al established an extended linear theory on FxE [12]. Maranganti *et al.* discussed a theoretical framework that can describe the size-dependent electromechanical coupling owing to strain or polarization gradients [13]. In addition, Hong *et al.* did a first attempt at calculating the longitudinal FxE coefficient for $SrTiO_3$ and $BaTiO_3$ from first-principles [14]. The first-principles theory of froze-ion FxE was developed very recently [15,16], which represents an important step in the direction of a full first-principles theory of FxE.

Despite the great advances in FxE study, there are few theoretical models and calculations addressing the FxE effect on the critical thickness of ferroelectric thin films below which the switchable spontaneous polarization vanishes. For epitaxial-grown thin films, the misfit strain between the film and substrate will relax as the film thickness increase. This will inevitably induce the strain gradient in the film and the FxE may have significant effect on the properties of ferroelectric films, such as polarization [9-11,17], dielectric constant [9-11], transition temperature [18], hysteresis loops [19] and critical thickness *etc*. The critical thickness is an important parameter for ferroelectric films below which the ferroelectric phenomena disappear and the films become paraelectric. However, so far very few works address the FxE effect on critical thickness. As the Landau-Ginsburg-Devonshire (LGD) theory has repeatedly been found to be very powerful in studying ferroelectric thin films, nanowires and other heterostructures down to the nanoscale [20-22], in this work, we use the LGD thermodynamic approach to investigate the FxE effect on the critical thickness of epitaxial $BaTiO_3$ ultrathin films. Our results show that the FxE increases the critical thickness and this effect is especially remarkable in the tension stressed films. In addition, we demonstrate that the film is still in a polar state even below the critical thickness due to the FxE effect, which is quite different from the conventional paraelectric state below the critical thickness without FxE.

## 2. Thermodynamic theory of FxE

We consider a *c*-phased ($P_1 = P_2 = 0$, and $P_3 = P \neq 0$) single-domain perovskite ferroelectric thin film epitaxially clamped onto a cubic substrate, with in-plane elastic stress resulted from lattice mismatch between the film and substrate, as shown in Fig. 1, where $h$ is the film thickness and $X$ the residual stress varied along the thickness direction.

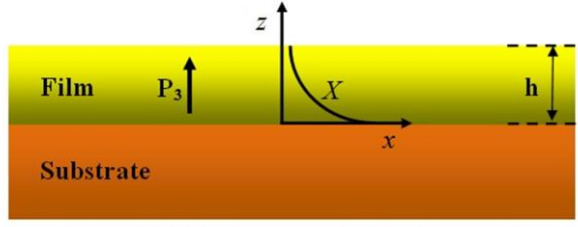

Fig. 1. Schematic of a film with in-plane mismatched stress in *x-z* plane.

The total free energy density of thin film $G$ can be expressed as follows

$$G = G_0 + G_1 + G_2, \qquad (1)$$

where $G_0$ is the free energy density of the paraelectric phase, $G_1$ the sum of electrostatic energy, elastic energy, depolarization field energy and surface energy, and $G_2$ the FxE energy density. $G_1$ and $G_2$ can be expressed as

$$G_1 = \frac{1}{h}\int_0^h \left\{ \frac{1}{2}\alpha_0 (T - T_{0\infty}) P^2 + \frac{1}{4}\beta P^4 + \frac{1}{6}r P^6 - \frac{1}{2}(s_{11} + s_{12}) X^2 - 2QXP^2 \right. \\ \left. + \frac{1}{2} K \left(\frac{dP}{dz}\right)^2 - \frac{1}{2} E_d P \right\} dz + \frac{K}{2}\left( \frac{P_i^2}{\delta_i} + \frac{P_s^2}{\delta_s} \right), \qquad (2)$$

$$G_2 = \frac{1}{h}\int_0^h \left( -\gamma \frac{dX}{dz} P - \eta X \frac{dP}{dz} \right) dz, \qquad (3)$$

where $h$ is the thickness of the film; the expansion coefficient $\alpha_0$, $\beta$ and $r$ are material parameters; $T_{0\infty}$ is the Curie-Weiss temperature (i.e. Curie temperature for second-order transition) of the bulk counterpart; $\gamma$ and $\eta$ FxE and converse FxE coupling coefficients; $Q$ electrostrictive coefficients; $s_{11}$ and $s_{12}$ elastic compliance coefficients; $P_i$ and $P_s$ the polarization of the interface ($z = 0$) and the surface ($z = h$) respectively; $\delta_i$ and $\delta_s$ the extrapolation length of the interface and the surface respectively; $\varepsilon$ the dielectric permittivity, and $E_d$ the depolarization field [23], i.e.

$$E_d = -1/\varepsilon \left( P - 1/h \int_0^h P dz \right) ; \qquad (4)$$

the residual stress $X$ is assumed to exponentially decrease from the film/substrate interface [24],

i.e.

$$X = X_0 \exp(-kz), \tag{5}$$

where $X_0$ is the maximum residual stress in the interface which is determined by the misfitted lattice constant [25]

$$X_0 = \frac{a_s - a_f}{a_s(s_{11} + s_{12})}, \tag{6}$$

where $a_s$ and $a_f$ are lattice parameters of the substrate and film materials. $k$ in Eq. (5) is thickness dependent and it is given by [24]: (in nm unit)

$$k(h) = k_0 - \zeta h = 3.925 \times 10^{-3} - 2.325 \times 10^{-6} h, \tag{7}$$

It is independent of temperature and has a dimension of (length)$^{-1}$. Eq. (7) indicates the strain gradient is larger in thinner films.

Variation of Eq. (1) yields the following Euler's equation:

$$K\frac{d^2 P}{dz^2} = \left[\alpha_0(T - T_{0\infty}) - 4QX_0 \exp(-kz) + \frac{1}{\varepsilon}\right]P + \beta P^3 + rP^5 - k(\eta - \gamma)X_0 \exp(-kz) - \frac{1}{h\varepsilon}\int_0^h P dz, \tag{8}$$

with the following boundary conditions:

$$\begin{cases} \dfrac{dP}{dz} = \dfrac{P}{\delta_i} & \text{when } z = 0 \\ \dfrac{dP}{dz} = -\dfrac{P}{\delta_s} & \text{when } z = h \end{cases}. \tag{9}$$

By using the finite-difference method, the polarization profile and the mean polarization of the film can be obtained.

## 3. Numerical calculation and discussion

In this paper, a BaTiO$_3$ nanofilm is taken as an example to show the effect of FxE on the critical thickness of ferroelectric films. According to recent experimental results by Ma and Cross [26], the value of FxE coefficient is about $\mu = 10$ μC/m at room temperature, and the value of the dielectric permittivity is $\varepsilon_r = 2360$. Then by using the relational expression obtained by Catalan *et*

*al.* [11]

$$\gamma - \eta = (s_{11} + s_{12})\varepsilon_0^{-1}\chi^{-1}\mu, \tag{10}$$

we have $\gamma - \eta = 2.69 \times 10^{-9}$ m$^3$C$^{-1}$, which is in agreement with the order of magnitude presented by Catalan *et al*. The properties of the BaTiO$_3$ thin films used in the simulations are presented as follows [23,27-29]: $T_{0\infty} = 383$ K, $\alpha_0 = 6.6 \times 10^5$ VmC$^{-1}$, $\beta = 14.4(T-448) \times 10^6$ Vm$^5$C$^{-3}$, $r = 39.6 \times 10^9$ Vm$^9$C$^{-5}$, $K = 0.9 \times 10^{-9}$ Vm$^3$C$^{-1}$, $Q = -0.043$ m$^4$C$^{-2}$, $\delta_i = \delta_s = \delta = 1$ nm, $s_{11} + s_{12} = 5.62 \times 10^{-12}$ m$^2$N$^{-1}$, $\gamma = 3.69 \times 10^{-9}$ m$^3$C$^{-1}$, $\eta = 1 \times 10^{-9}$ m$^3$C$^{-1}$.

For the convenience of comparison, we introduce the relative mean polarization $\bar{p} = \bar{P}/P_\infty$, where $\bar{P}$ is the mean value of polarization along the thickness direction of the film, $P_\infty = 0.27$ Cm$^{-2}$ is the polarization value of the bulk BaTiO$_3$ material. The size effect of BaTiO$_3$ films from this model agrees well with available experimental results, showing that the approach is effective and accurate at nanoscale (see appendix). The positive and negative polarization represents the polarization orientating towards the surface and interface, respectively. Furthermore, we consider the interface stresses $X_0 = -2.0$ GPa and $X_0 = 0.3$ GPa in the calculation which is the same as that in the experiments [30,31].

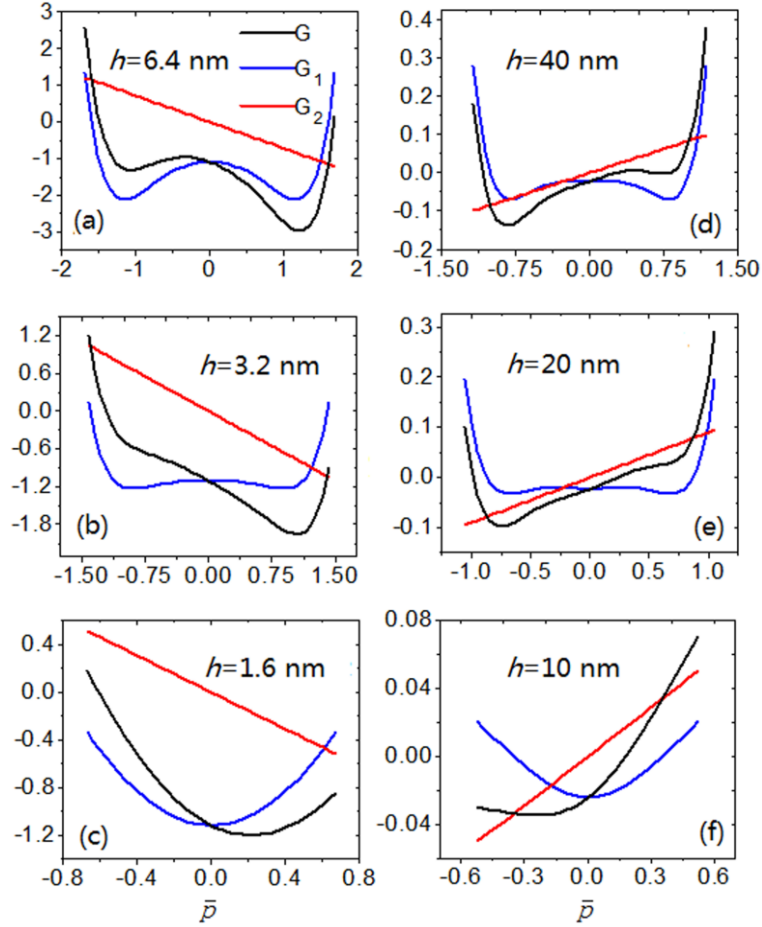

Fig. 2. Energy profile for different thickness films with misfit stress $X_0 = -2.0$ GPa (left column) and $X_0 = 0.3$ GPa (right column) at room temperature. Free energy of paraelectric state $G_0$ is set to zero. Energy density unit: $10^7 \text{Nm}^{-2}$.

Figure 2 shows the free energy profile for different thickness films under compressive and tensile stresses. Taking the compressively strained films (Figs.2a-2c) for an example, we can see that the $G_1$-$\bar{p}$ profile (blue curves) is the well-know ferroelectric double-well potential for thick film and the potential barrier decreases as thickness becomes thinner (Figs.2a-2b), and then $G_1$-$\bar{p}$ profile turns to a single-well potential with energy minimum at $\bar{p} = 0$ (Fig.2c). We call this critical thickness as $h_{c1}$ below which the ferroelectric disappears without considering FxE. The FxE energy $G_2$ (red curves) decreases as polarization increases from negative to positive value for films with different thickness.

However, the total free energy $G = G_1 + G_2$ ($G_0$ is set to zero) shows some interesting features. The $G$-$\bar{p}$ profile (black curve) shows asymmetrical double-well for thick film due to FxE effect (Fig.2a). This asymmetrical double-well is also proposed in recent flexoelectric work based on the measurement [32]. The energy curve has two non-equivalent local energy minima, indicating two non-equivalent ferroelectric states (imprint behavior [2]). The most stable state is the positive polarization for a compressive film. As film thickness decreases (Fig.2b), the total energy shows only one local minimum at positive polarization, suggesting the ferroelectricity disappears and the film becomes a uni-polar material due to the FxE effect. We call this critical thickness as $h_{c2}$ below which the ferroelectricity disappears and the film becomes uni-polar material considering FxE. This is quite different from the situation without FxE ($G_1$), in which the same-thickness film is still ferroelectrics in theory.

Figs.2a-2c shows that the size-induced phase transition is different for the cases with and without FxE effect. Without FxE effect, as thickness reduces to critical thickness $h_{c1}$, the ferroelectric disappears and film becomes paraelectrics below $h_{c1}$. However, if considering the FxE effect, the ferroelectric disappear and the film becomes a uni-polar material below $h_{c2}$. For tensile strain, the similar behaviours are observed as shown in Figs.2d-2f. Note that the preferred ferroelectric state for tension strained films is the polarization orientating towards the interface, i.e., the negative polarization state.

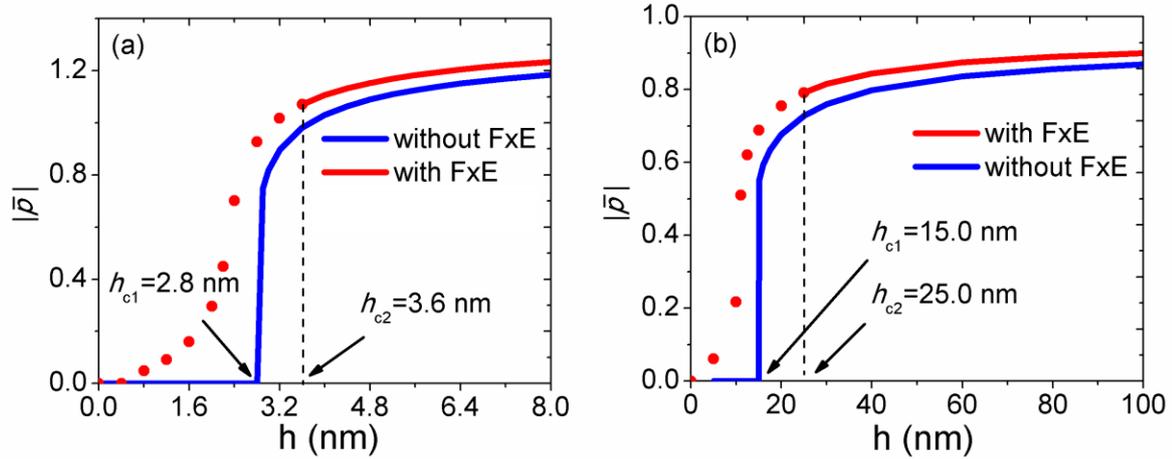

Fig. 3. Normalized mean polarization $|\bar{p}|$ of the films with different thicknesses at room temperature. (a) compressive stress, $X_0 = -2.0 \text{ GPa}$; (b) tensile stress, $X_0 = 0.3 \text{ GPa}$. The red dots indicate the single polar state.

Figure 3 shows the mean polarization in BaTiO$_3$ films with and without FxE. The FxE enhances the polarization for films with the interface misfit stress $X_0$ whether compressive or tensile. However, this enhanced polarization does not stabilize ferroelectricity in thinner films but suspend it. Fig.3a shows that the critical thickness $h_{c1}$ (without FxE) is 2.8 nm (seven unit-cells) for a film with interface misfit compressive stress $X_0 = -2.0 \text{ GPa}$, but if considering FxE effect, the critical thickness $h_{c2}$ increases to 3.6 nm (nine unit-cells), below which the BaTiO$_3$ film becomes a uni-polar material due to the FxE as Figs.2b-2c show. The reason for the critical thickness increase is that the stress gradient in films breaks the inversion symmetry, its coupling with polarization induces an asymmetrical double-well potential and makes one well deeper while the other well shallower. As film thickness reduces to $h_{c2}$, the shallow well potential disappears and the BaTiO$_3$ film turns to a uni-polar state. The polarization of this non-ferroelectric BaTiO$_3$ film decreases continuously and disappears as film thickness approaches zero (red dots in Fig.3a). The similar increase of critical thickness for tensile strained films is shown in Fig.3b, with the critical thickness $h_{c1} = 15.0$ nm and $h_{c2} = 25.0$ nm, respectively.

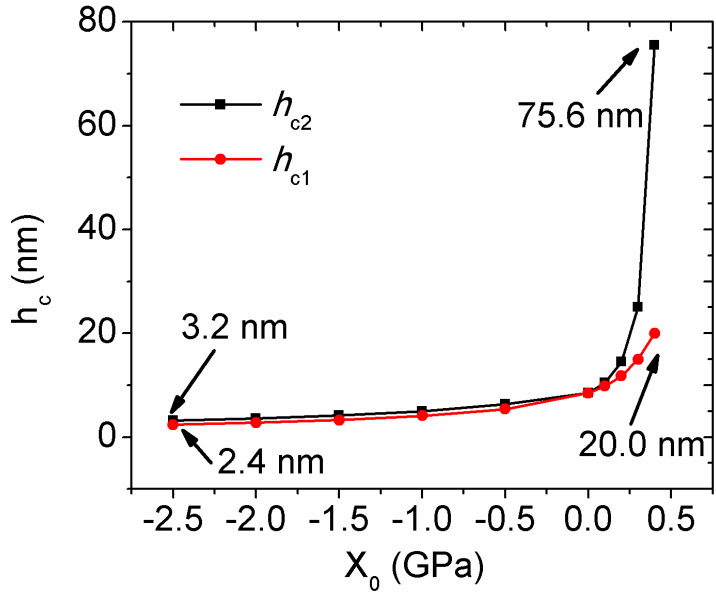

Fig. 4. Influence of interface misfit stress $X_0$ on critical thickness.

Figure 4 shows the influence of the misfit stress on the critical thickness of thin films. It can be seen from Fig. 4 that the compressive stress reduces the critical thicknesses while the tensile stress increases them. The critical thickness $h_{c2}$ is larger than $h_{c1}$, indicating that FxE can increase the critical thickness for both compressive and tensile stressed films. It also shows that $h_{c2}$ is only slightly larger than $h_{c1}$ under compressive misfit stress with the difference less than 1.0 nm, for instance, $h_{c1} = 2.4$ nm and $h_{c2} = 3.2$ nm for $X_0 = -2.5$ GPa. However, the effect of FxE on critical thickness is significant when the film is under tension stress. For instance, under $X_0 = 0.4$ GPa, $h_{c1}$ is 20.0 nm while $h_{c2}$ increases sharply to 75.6 nm, the FxE enhances critical thickness nearly four times of that without FxE under such a low level tension stress.

The significant difference between the inhomogeneous compressive and the tensile stresses in affecting the critical thickness for ferroelectricity is actually due to the electrostriction effect. As can be seen from Eq. (2-3), the change of the signs of stress $X$ and polarization $P$ simultaneously can change the free energy only through the electrostriction term $-2QXP^2$ while all the other terms remain unchanged, which results in the lower barrier height of $G_1$ under tensile stress than that under compressive stress (for films with the same thickness and same magnitude of interface tensile or compressive misfit stresses). For that reason, the potential well profile of $G_1$ can be more easily affected by the FxE contribution $G_2$ within a particular level ($X_0$ ranging from -2.5 to 0.5 GPa), i.e., the ferroelectricity of films under tensile stress can be more easily affected by the FxE, resulting in the obviously increased critical thickness compared with that in compressively strained films (see Apendix for the details). Therefore, large tensile misfit stress gradient should be avoided to prevent the degradation of device performance in ferroelectric thin films.

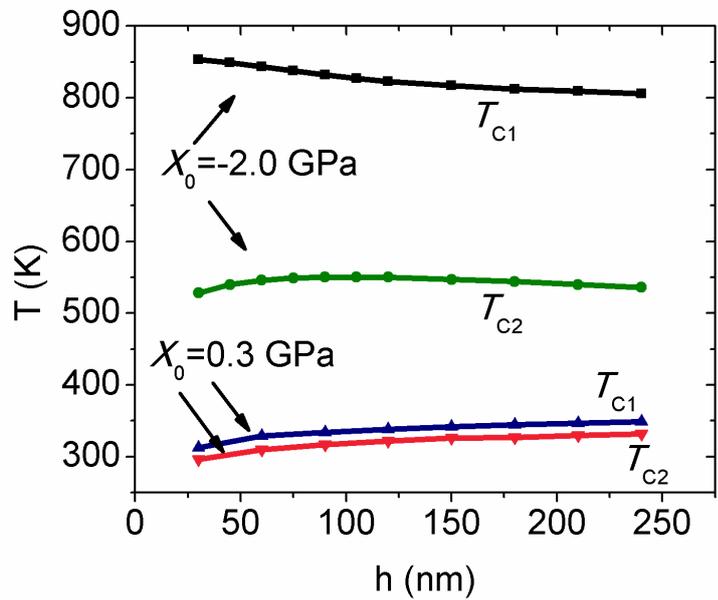

Fig. 5. FxE induces the decrease of critical temperature for ferroelectricity. $T_{C1}$ and $T_{C2}$ are the critical temperatures for $BaTiO_3$ film with and without FxE, respectively.

In addition, the increase in critical thickness for ferroelectricity is, in energy terms, equivalent to a decrease in critical temperature. Fig. 5 shows that FxE induces the decrease of critical temperature above which the ferroelectricity disappears. From this figure we can see that the large compressive stress ($X_0 = -2.0$ GPa) improves the critical temperature $T_{C1}$ from bulk value ($T_{0\infty} = 383$ K) to ~850 K without FxE considered. However, if FxE is taken into account, the critical temperature $T_{C2}$ reduces significantly to ~550 K. The tensile stress gradient also induces the decrease of critical temperatures, but the reduction is less than that of compressive stress gradient. This FxE inducing decrease of critical temperature agrees with the predictions in Ref. 11.

## 4. Conclusions

In summary, we used the LGD phenomenological theory to study the critical thickness of epitaxial BaTiO$_3$ thin films, taking both the FxE and the surface effect into account. The total free energy profiles under compressive and tensile misfit stresses are investigated and the increase of critical thickness is found in BaTiO$_3$ films due to FxE. This increase is remarkable in tensile films but is trivial in compressive films. The differences between the tensile and the compressive inhomogeneous stresses in affecting the critical thickness is because of the electrostriction induced changes in the potential well depth and further more the corresponding changes in the ability to retain ferroelectricity in the presence of FxE. Our results show that large tension stress gradient should be avoided in order to prevent the degradation of device performance in ferroelectric thin films. In addition, the films are still in a uni-polar state even below the critical thickness due to FxE, which is quite different from those being paraelectric state under critical thickness without FxE.

It should be noted that in this work we assume an exponential decay of stress inside the film along the thickness direction according to the Ref. 24, however, the function form may not describe the real stress distribution inside thin films synthesized under different conditions very well in nano scale. Especially, there are unlikely to be any strain gradients in ultrathin films that are too thin to overcome the Matthews-Blackeslee barrier [33] for the onset of misfit dislocations which play an essential role in strain relaxation of epitaxial films. In this case, there is no strain gradient in the ultrathin films and FxE effect disappears, therefore only strain effect needs to be considered in the model.

**Acknowledgments**

The authors are grateful for the support by National Natural Science Foundation of China under grants 11090330, 11090331 and 11072003. Support by the National Basic Research Program of China (G2010CB832701) is also acknowledged.

**Appendix**

The compare between the experimental data from references [30, 31] and our simulation is presented in Fig. A.1, which shows that the model is effective and accurate at nanoscale.

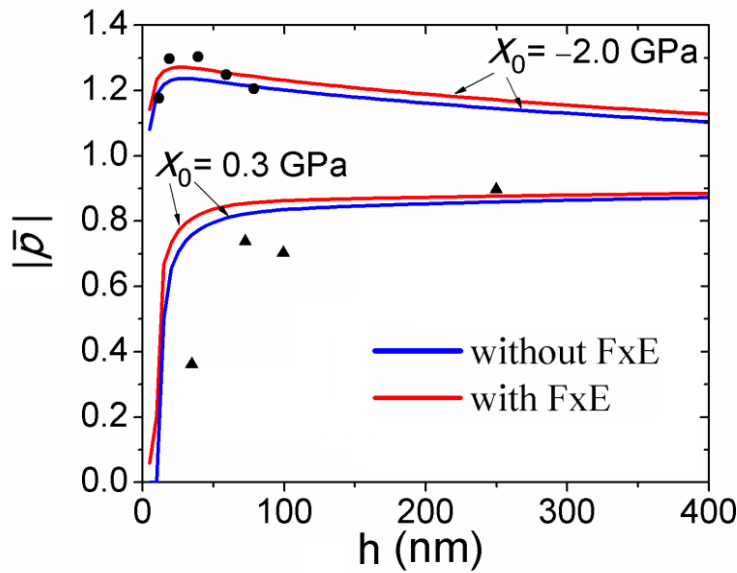

Fig. A1. Normalized mean polarization of the films with different thicknesses in absence of external electric field at room temperature. The experimental data are from reference [30] (solid dots for $X_0$=-2.0 GPa) and reference [31] (solid triangles for $X_0$=0.3 GPa)

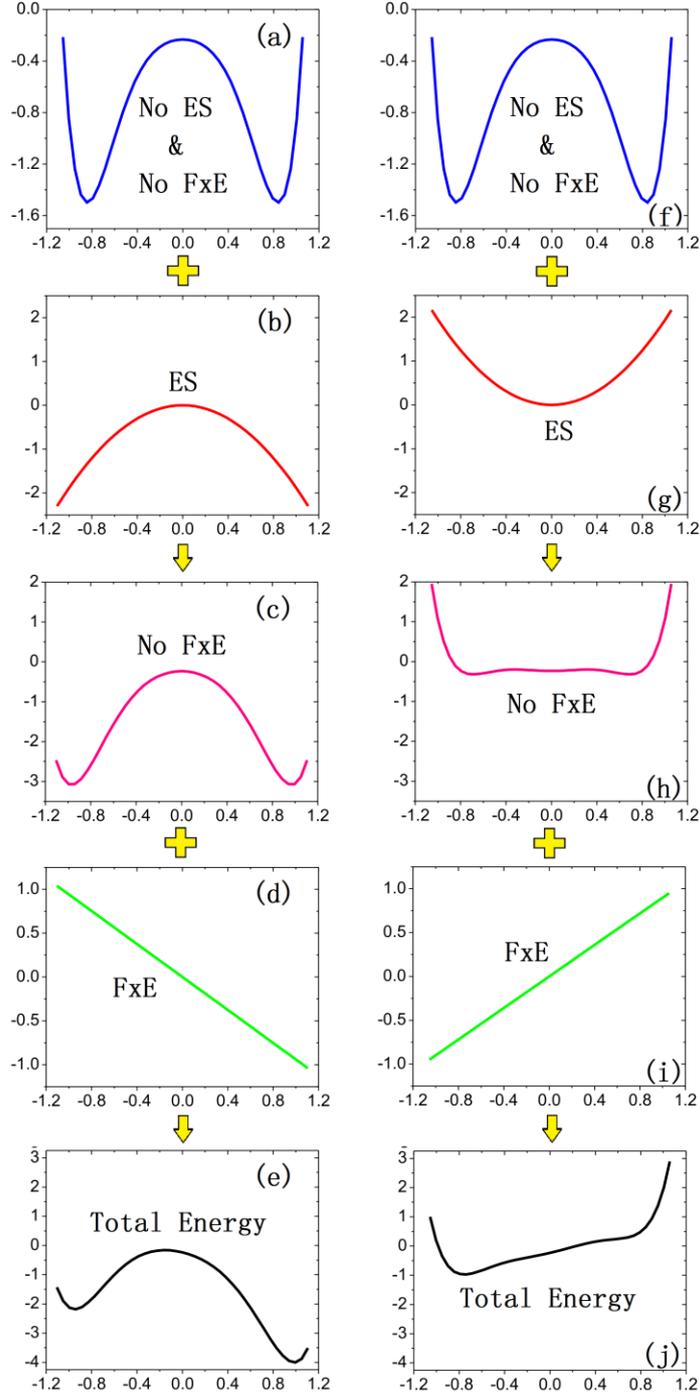

Fig. A2. The competition between the ES (electrostriction) and the FxE (flexoelectricity) in affecting the potential profiles in films with thickness $h$=20 nm under compressive and tensile interface misfit stresses $X_0$=-0.3 GPa (left column) and $X_0$=0.3 GPa (right column), respectively. Free energy of paraelectric state $G_0$ is set to zero. Energy density unit: $10^6 Nm^{-2}$.

As Figs. A2a, A2f show, the energy profiles is identical under the same magnitude compressive or tensile stresses without ES and FxE. However, the superposition of the ES energy, which is opposite in sign for compressive and tensile stresses, changes the potential well depth or the potential barrier height between the ferroelectric and the paraelectric states (as Figs. A2c, A2h show), and then, the shallower well or the lower barrier under tensile stress can be easier affected by the FxE energy just as Figs. A2e, A2j show, and the result is that the total energy profile is still double well for compressively strained films while already single well for tensile strained films. In brief, the ES makes the asymmetry between the compressive and the tensile inhomogeneous stresses in affecting the critical thickness through the FxE.

**References**

[1] S. M. Kogan, Sov. Phys. Sol. Stat. **5**, 10 (1964).

[2] A. Gruverman, B. J. Rodriguez, R. J. Nemanich, A. I. Kingon, J. Appl. Phys. **92**, 2734 (2002).

[3] A. Gruverman, B. J. Rodriguez, A. I. Kingon, R. J. Nemanich, A. K. Tagantsev, J. S. Cross, M. Tsukada, Appl. Phys. Lett. **83**, 728 (2003).

[4] H. X. Cao, V. C. Lo, Z. Y. Li, Solid State Commun. **138**, 404 (2006).

[5] R. Maranganti, P. Sharma, Phys. Rev. B **80**, 5 (2009).

[6] A. K. Tagantsev, Phase Transit. **35**, 119 (1991).

[7] L. E. Cross, J. Mater. Sci. **41**, 53 (2006).

[8] W. Ma, L. E. Cross, Appl. Phys. Lett. **78**, 2920 (2001); **79**, 4420 (2001); **81**, 3340 (2002); **82**, 3293 (2003); **86**, 072905 (2005).

[9] P. Zubko, G. Catalan, A. Buckley, P. R. L. Welche, J. F. Scott. Phys. Rev. Lett. **99**, 167601 (2007).

[10] G. Catalan, B. Noheda, J. McAneney, L. J. Sinnamon, J. M. Gregg, Phys. Rev. B **72**, 020102(R) (2005).

[11] G. Catalan, L. J. Sinnamon, J. M. Gregg. J. Phys-condens. Matter **16**, 13 (2004).

[12] S. P. Shen, S. L. Hu, J. Mech. Phys. Solids **58**, 5 (2010).

[13] R. Maranganti, N. D. Sharma, P. Sharma, Phys. Rev. B **74**, 014110 (2006).

[14] J. W. Hong, G. Catalan, J. F. Scott, E. Artacho, J. Phys.: Condens. Matter **22**, 112201 (2010).

[15] R. Resta, Phys. Rev. Lett. **105**, 127601 (2010).

[16] J. W. Hong, D. Vanderbilt, Phys. Rev. B **84**, 180101 (2011).

[17] G. Catalan, A. Lubk, A. H. G. Vlooswijk, E. Snoeck, C. Magen, A. Janssens, G. Rispens, G. Rijnders, D. H. A. Blank, B. Noheda, Nature Mater. **10**, 963–967 (2011).

[18] E. A. Eliseev, A. N. Morozovska, M. D. Glinchuk, R. Blinc, Phys. Rev. B **79**, 165433 (2009).

[19] D. Lee, A. Yoon, S. Y. Jang, J.-G. Yoon, J.-S. Chung, M. Kim, J. F. Scott, T. W. Noh, Phys. Rev. Lett. **107**, 057602 (2011).

[20] C. H. Ahn, K. M. Rabe, J. M. Triscone, Science **303**, 488 (2004).

[21] J. W. Hong, D. N. Fang, Appl. Phys. Lett. **92**, 012906 (2008)

[22] J. W. Hong, D. N. Fang, J. Appl. Phys. **104**, 064118 (2008)

[23] G. Liu, C. W. Nan, J. Phys. D: Appl. Phys. **38**, 4 (2005).

[24] H. J. Kim, S. H. Oh, H. M. Jang, Appl. Phys. Lett. **75**, 3195 (1999).

[25] N. A. Pertsev, A. G. Zembilgotov, A. K. Tagantsev, Phys. Rev. Lett. **80**, 1988 (1998).

[26] W. H. Ma, L. E. Cross, Appl. Phys. Lett. **88**, 232902 (2006).

[27] N. A. Pertsev, A. G. Zembilgotov, A. K. Tagantsev, Phys. Rev. Lett. **80**, 9 (1998).

[28] A. N. Morozovska, E. A. Eliseev, M. D. Glinchuk, Phys. Rev. B **73**, 214106 (2006).

[29] X. S. Wang, C. L. Wang, W. L. Zhong, D. R. Tilley, Solid State Commun. **121**, 111 (2002).

[30] N. Yanase, K. Abe, N. Fukushima, T. Kawakubo, Japan. J. Appl. Phys. **38**, 5305 (1999).

[31] G. F. Huang, S. Berger, J. Appl. Phys. **93**, 2855 (2003).

[32] H. Lu, C.-W. Bark, D. Esque de los Ojos, J. Alcala, C.-B. Eom, G. Catalan, A. Gruverman, Science **336**, 59 (2012).

[33] J. W. Matthews, W. E. Blakeslee, J. Cryst. Growth **27**, 118 (1974); **29**, 273 (1975).